\documentclass[trackchanges,astrosymb,longbib]{aastex701-modified}
\usepackage{xcolor}
\usepackage{fancyhdr}
\newenvironment{tabitemize}{%
 \setlength{\leftmargin}{-3pt}
 \begin{itemize}}
 {\end{itemize}}

\begin{document}

\title{The MegaWave Radio Surveyor}

\author[orcid=0009-0003-8984-388X,gname='T. Joseph W.',sname=Lazio]{T.~Joseph~W.~Lazio (\textup{Michigan})}
 \email{}

\author[gname='Evgenya L.',sname=Shkolnik]{Evgenya Shkolnik (ASU)}
 \email{}
 
\author[gname=James, sname=Aguirre]{James Aguirre (\textup{Pennsylvania})}
 \email{}
 
\author[gname='Stuart D.', sname=Bale]{Stuart D.~Bale (\textup{UC~Berkeley})}
 \email{}
 
\author[gname=Judd, sname=Bowman]{Judd Bowman (ASU)}
 \email{}
 
\author[gname=Ruby, sname=Byrne]{Ruby Byrne (\textup{Caltech})}
 \email{rbyrne@caltech.edu}

\author[gname=Joseph, sname=Callingham]{Joseph Callingham (ASTRON)}
  \email{}
  
\author[gname='Tracy E.', sname=Clarke]{Tracy E.~Clarke (NRL)}
    \email{}
    
\author[gname=Ivey,sname=Davis]{Ivey Davis (ASTRON)}
 \email{}

\author[gname=Tim, sname=Dolch]{Tim Dolch (UNM)}
  \email{}
  
\author[gname=Peter, sname=Driscoll]{Peter Driscoll (\textup{Carnegie Science})}
 \email{}

\author[gname=Anastasia, sname=Fialkov]{Anastasia Fialkov (\textup{Cambridge})}
 \email{}
    
\author[gname=Steven, sname=Furlanetto]{Steven Furlanetto (UCLA)}
 \email{}
   
\author[gname=Simona, sname=Giacintucci]{Simona Giacintucci (NRL)}
 \email{}

\author[gname=Ravit, sname=Helled]{Ravit Helled (\textup{Univ.~Zurich})}
 \email{r.helled@gmail.com}

\author[gname=Jacqueline, sname=Hewitt]{Jacqueline Hewitt (MIT)}
 \email{}
 
\author[gname=Phil, sname=Hopkins]{Phil Hopkins (\textup{Caltech})}
 \email{}
 
\author[gname=Andrea, sname=Isella]{Andrea Isella (\textup{Rice})}
 \email{}
 
\author[gname=Daniel, sname=Jacobs]{Daniel Jacobs (ASU)}
 \email{}

\author[gname=Eloy, sname='de Lera Acedo']{Eloy de~Lera Acedo (\textup{Cambridge})}
 \email{}

\author[gname='Melodie M.', sname=Kao]{Melodie M. Kao (\textup{Lowell Observatory})}
 \email{}

\author[gname=Mary, sname=Knapp]{Mary Knapp (\textup{MIT/Haystack Observatory})}
 \email{}

\author[gname='L. V. E.', sname=Koopmans]{L.{}V.{}E.~Koopmans (\textup{Kapteyn Astronomical Institute})}
 \email{}
 
\author[gname=Nicholas, sname=Kern]{Nicholas Kern (\textup{Michigan})}
 \email{}
 
\author[gname=Jasmina, sname='Lazendic-Galloway']{Jasmina Lazendic-Galloway (\textup{Eindhoven University of Technology})}
 \email{}

 \author[gname='Susan T.', sname=Lepri]{Susan T.~Lepri (\textup{Michigan})}
 \email{}

\author[gname='R. O. Parke', sname=Loyd]{R.~O.~Parke Loyd (\textup{Eureka Scientific})}
 \email{}
 
\author[gname='James P.', sname=Lux]{James P.~Lux (\textup{\hbox{JPL}, Caltech})}
 \email{}
 
\author[gname=James, sname=Mason]{James P. Mason (APL/JHU)}
 \email{}
 
\author[gname=Raul, sname=Monsalve]{Raul Monsalve (\textup{Universidad Andr\'es Bello})}
 \email{}
 
\author[gname=Miguel, sname=Morales]{Miguel Morales (\textup{Univ.~Washington})}
 \email{}
 
\author[gname=Julian, sname=Munoz]{Julian B.~Mu{\~n}oz (\textup{UT~Austin})}
  \email{}
  
\author[gname=Rachel, sname=Osten]{Rachel Osten (\textup{Johns Hopkins; STScI})}
 \email{}
 
\author[gname=Sebastian, sname=Pineda]{J.~Sebastian Pineda (\textup{CU Boulder})}
   \email{}

\author[gname=Jonathan, sname=Pober]{Jonathan Pober (\textup{Brown})}
   \email{}
   
\author[gname='Sam B.', sname=Ponnada]{Sam B.~Ponnada (\textup{Chalmers})}
 \email{}

\author[gname='Leslie A.', sname=Rogers]{Leslie A.~Rogers (\textup{Chicago})}
   \email{}
   
\author[gname=Saurabh, sname=Singh]{Saurabh Singh (\textup{Raman Research Institute})}
   \email{}
   
\author[gname='Jake D.', sname=Turner]{Jake D.~Turner (\textup{Cornell})}
   \email{}
   
\author[gname=Jackie, sname=Villadsen]{Jackie Villadsen (\textup{Bucknell})}
   \email{}
   
\author[gname=Philippe, sname=Zarka]{Philippe Zarka (\textup{Observatoire de~Paris})}
   \email{}
   
\author[gname=John, sname=ZuHorne]{John ZuHorne (\textup{Harvard-Smithsonian CfA})}
   \email{}
   
\author[gname=Ellen, sname=Zweibel]{Ellen Zweibel (\textup{Univ.~Wisconsin, Madison})}
   \email{}

\begin{abstract}
Several Decadal-level questions in astrophysics, planetary science, astrobiology, and cosmology can be addressed only at low radio frequencies that are inaccessible from Earth.
The MegaWave Radio Surveyor would open this largely-unexplored region of the electromagnetic spectrum with a distributed space-based interferometer to 
(1)~Track space weather events generated by other stars (exospace weather);
(2)~Detect the magnetically-generated radio emission from exoplanets to probe their interiors and assess the magnetic shielding of their atmospheres; 
(3)~Probe directly the Universe's evolution during the Dark Ages via the highly-redshifted hyperfine line of neutral hydrogen~\ion{H}{1};
and
(4)~Assess the role of cosmic rays and magnetic fields in shaping the structures of galaxies and clusters of galaxies in the cosmic web.
Developed in the context of NASA's Astrophysics Strategic Technology \& Research Accelerator (ASTRA) Initiative, the MegaWave Radio Surveyor’s science objectives directly respond to priorities identified in the \textit{Pathways to Discovery} Decadal Survey and three other National Academies studies, and the MegaWave Radio Surveyor would serve as a Formative Era concept in the \textit{Enduring Quests, Daring Visions} roadmap.

For the first time, the capabilities required to realize such an observatory are within reach, enabled by ongoing developments in U.{}S.\ space industries including the standardization of small spacecraft, lower launch costs, advances in onboard processing, and communications.
The MegaWave Radio Surveyor would offer a versatile and scalable architecture capable of highly sensitive observations below approximately 45\,MHz, unprecedented angular resolution for these frequencies, multiple simultaneous science investigations, and enhanced resilience through its multi-element design.
The concept leverages NASA's experience with the development and operations of the Sun Radio Interferometer Space Experiment (SunRISE), the Star-Planet Activity Research CubeSat (SPARCS), and the Lunar Surface Electromagnetics Experiment (LuSEE-Night).
The MegaWave Radio Surveyor could leverage multiple elements of the Artemis/Moon-to-Mars program including access to and beyond cislunar space and communications-navigation capabilities in cislunar space; and 
the concept could offer opportunities to infuse new modes for mission operations involving higher levels of autonomy and artificial intelligence/machine learning (AI/ML).

Beyond its specific science objectives, by opening one of the last unexplored windows in the electromagnetic spectrum and pioneering distributed space interferometry at unprecedented scales, the MegaWave Radio Surveyor would establish a transformational new capability for NASA and the global astronomical community.

\end{abstract}


\clearpage
\newpage

\begin{figure}[h!]  
  \includegraphics[width=\textwidth]{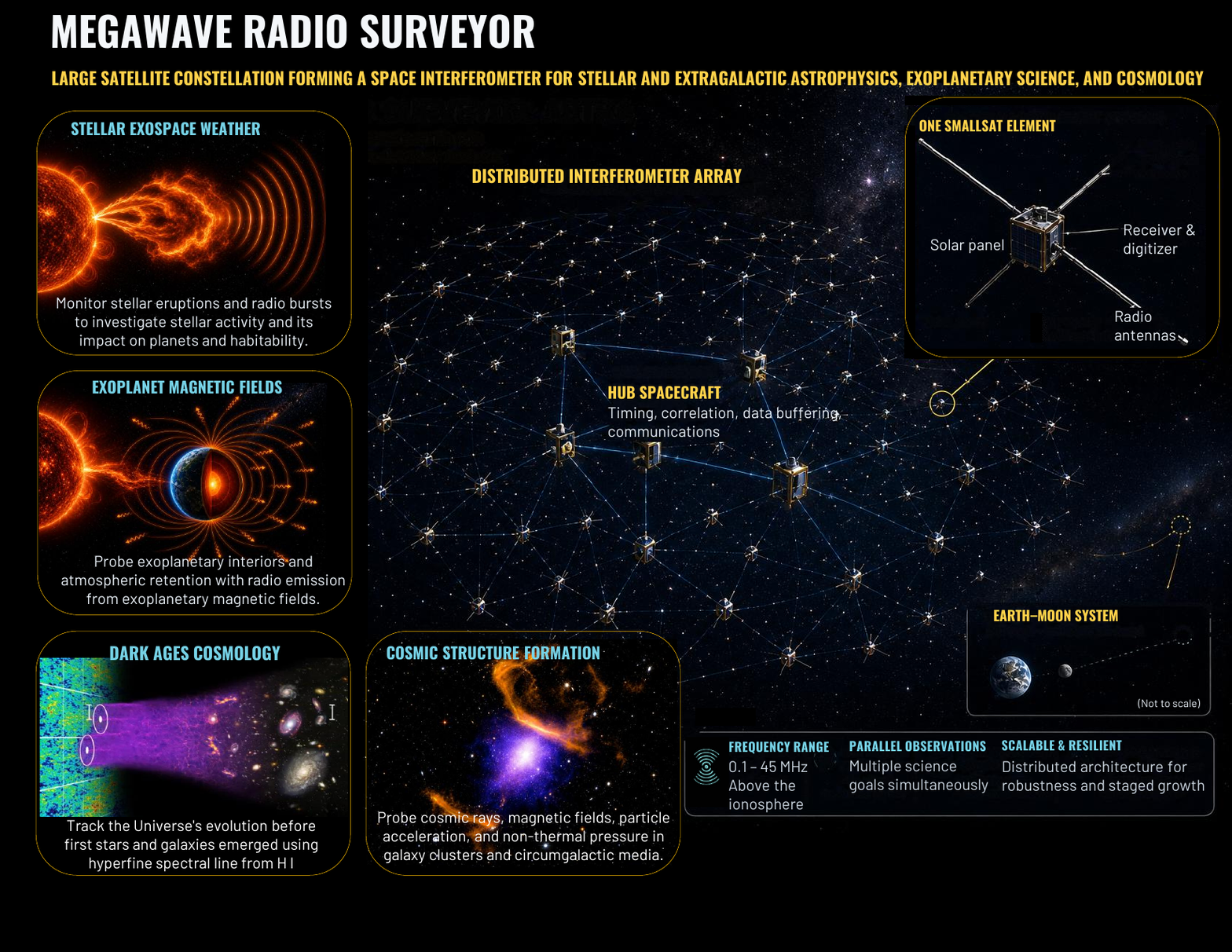}
  \label{fig:intro}
\end{figure}

\setcounter{figure}{0}

\vspace*{1pt}
\bigskip
\section{Science Investigation}\label{sec:science}

Decadal-level science questions spanning astrophysics, exoplanetary science, astrobiology, and cosmology require observations at radio frequencies blocked by Earth's ionosphere.
However, radio waves propagating through the ionosphere exhibit measurable phase changes at frequencies even well above~1\,GHz, and, at frequencies below about~100\,MHz, there is increasing absorption. 
These detrimental effects are reflected in the fact that the International Telecommunications Union (ITU) provides no recommendations for radio astronomical observations below~13.6\,MHz \citep{hra2013}.
The recognition that a space-based radio telescope could escape the effects of the Earth's ionosphere is long-standing \citep{aghl58,1963rat..conf...54J,1969Sci...166..775P} and has led to a number of initial descriptions and specific mission proposals \citep[e.g.,][]{wjs+88,bbd+97a,bbd+97b,2000GMS...119....1K}, notably including the (NASA MIDEX) Astronomical Low Frequency Array \citep[\hbox{ALFA},][]{alfa}, (ESA concept) Nanosatellites pour un Observatoire Interf{\'e}rom{\'e}trique Radio dans l'Espace/Nanosatellites for a Radio Interferometer Observatory in Space \citep[\hbox{NOIRE},][]{NOIRE}, and (Chinese-led) Discovering the Sky at the Longest Wavelengths \citep[{DSL} or ``Hongmeng,''][]{2023ChJSS..43...43C}.

What has changed from many of these previous mission concepts is both a sharper set of science questions and an increasing U.{}S.\ space industry capable of delivering such a mission.
In the following sub-sections, we describe motivating science cases for the MegaWave Radio Surveyor, a concept developed in the context of NASA's Astrophysics Strategic Technology \& Research Accelerator (ASTRA) Initiative, casting them in terms of Science Questions (Table~1) from the \textit{Pathways to Discovery} Decadal Survey.
We state specific Science Objectives for the MegaWave Radio Surveyor that flow from these science cases.
We also describe links from these science investigations to other National Academies' reports.
Subsequent sections outline the implementation of the MegaWave Radio Surveyor as an interferometer (\S\ref{sec:instrument}) and the larger mission implementation (\S\ref{sec:mission}), but, consistent with the spirit of the ASTRA Initiative, we focus on high-level aspects rather than presenting a so-called ``point design.'' 

While the surface of the Moon long has been identified as a potential location for some of these observations, the concept of a free-flying space-based interferometer also is consistent with findings from community workshops that ``observations from free space (\ldots) offer the most promise for significant progress in broad areas of astrophysics'' \citep{2008LPICo1415.2155L}.
Finally, beyond its specific science objectives, by opening one of the last unexplored windows in the electromagnetic spectrum, the MegaWave Radio Surveyor would establish a transformational new capability for NASA and the broader Astrophysics community.

\subsection{Decadal Science Question: How do the Sun and other stars create space weather?}\label{sec:exospace}

Solar System planets are immersed in space weather---a steady solar wind and full-spectrum radiation bath punctuated by impulsive, transient events \citep{2021LRSP...18....4T}.
Transient space weather events---solar flares, coronal mass ejections (CMEs), and energetic particles (accelerated to trans-relativistic speeds)---result from the rapid conversion of magnetic energy into kinetic energy.
Space weather represents a fundamental aspect of the Sun's evolution
through which it loses mass and angular momentum \citep{KISS_CME}.
Space weather also transforms planetary environments, alters
atmospheric chemistry, drives atmospheric escape, and influences long-term habitability.
Mars' atmosphere is a dramatic example of the effects of space weather, as the combination of the steady solar wind and transient events, notably CMEs, likely played a substantial role to its erosion over the planets' history, as illustrated by observations from the Mars Atmosphere and Volatile Evolution (MAVEN) mission \citep{jgl+15,2018Icar..315..146J,2023JGRA..12830884J}.

\begin{figure}[bt]
  \centering
  \label{fig:exospace}
  \makebox[0.50\textwidth][c]{%
    \includegraphics[width=0.50\textwidth]{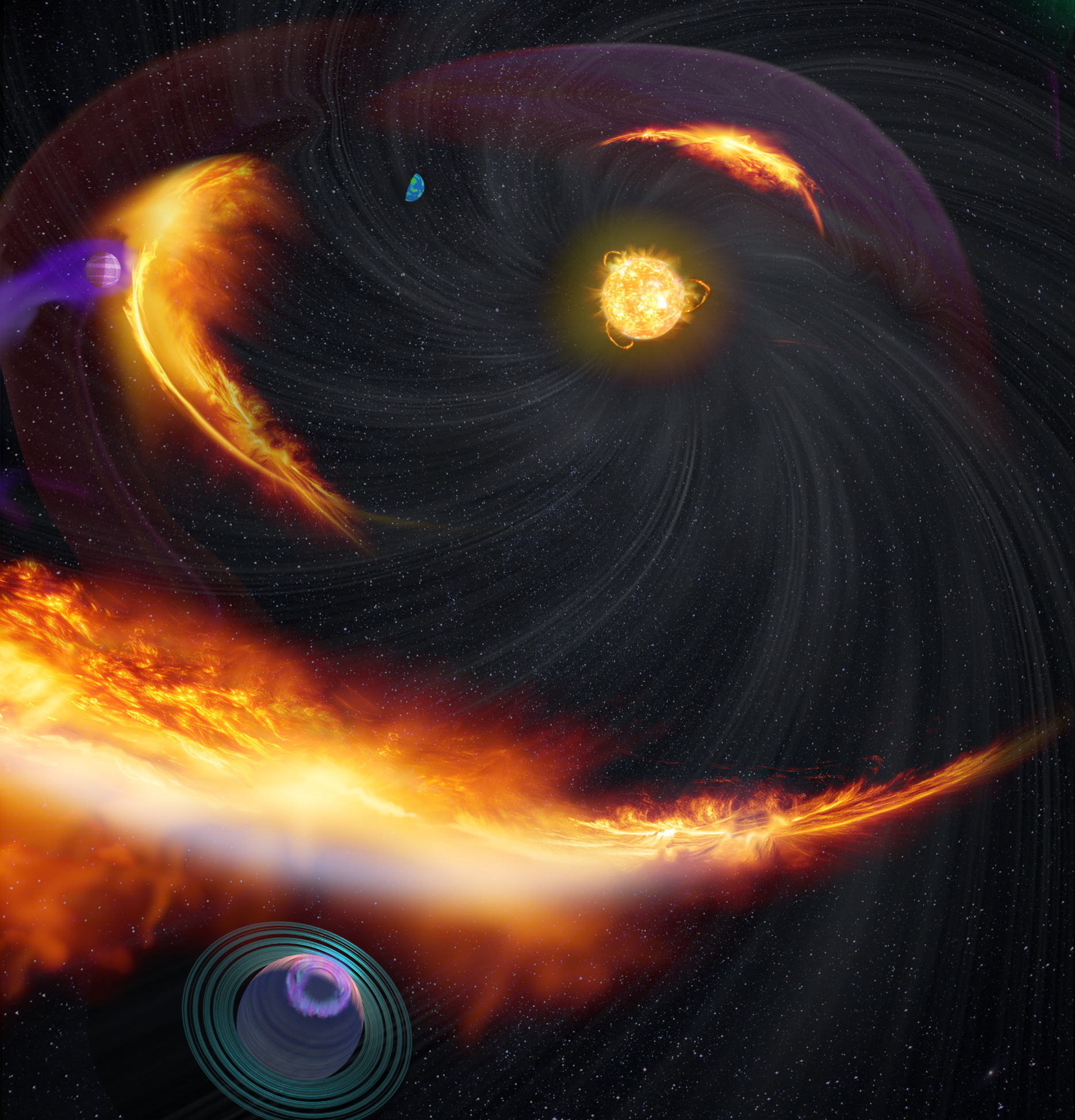}}\hfil%
  \parbox[b]{0.45\textwidth}{%
    \caption{All stars generate \emph{exospace weather}, a combination of a steady stellar wind and energetic, transient events, that affects both their evolution via a steady mass loss and potentially affects any planets that they host.  The MegaWave Radio Surveyor would use transient radio bursts occurring at frequencies below about~15\,MHz that result from (1)~electron beams generated by stellar flares to measure stellar wind densities (so-called Type~III bursts), and~(2)~energetic electrons accelerated at shock waves driven by stellar coronal mass ejections (so-called Type~II bursts). These observations of the radio emission from energetic particles are critical to assess the potential effects of exospace weather on planetary systems. (Credit: Keck Institute for Space Studies/Chuck Carter)}}
\end{figure}

While there is evidence that other stars generate equivalent \emph{exospace weather} \citep{1991ARAandA..29..275H,2004LRSP....1....2W,2014ASTRP...1...43L,2021ApJ...915...37W}, estimates of stellar mass and angular momentum loss have significant uncertainties, and some models of stellar coronae even predict that sufficiently strong overlying stellar magnetic fields may suppress the launching of CMEs \citep{2018ApJ...862...93A,2026arXiv260416156C}.
Planets around such stars would be exposed to less extreme exospace weather than otherwise might be expected based on usual measures of stellar activity.  
While stellar electromagnetic emissions are observed regularly, a model-independent, broadly applicable measure of stellar particle fluxes or winds is lacking, especially of high-energy particles that can transform planetary atmospheres, leaving the full extent of exospace weather and its effects on their planets unknown \citep[Figure~1,][]{KISS_CME}.

In contrast to the more than a quarter-century of \textit{in situ} and
multi-spacecraft measurements of the Sun's space weather, characterizing
exospace weather presents obvious challenges.
Moreover, the Sun is a single G~dwarf star, observed at a specific
time in its evolution.
Observations of exospace weather for stars with a diverse range of masses and ages would provide fundamental information
about stellar properties and planetary environments, a task that is crucial to characterizing
exoplanet habitability in the Habitable Worlds Observatory (HWO) era,
while simultaneously placing the Sun as a star in a galactic context.

Solar space weather processes produce plasma emissions---radio frequency waves are generated when energetic electrons excite waves in the solar wind and at a frequency determined by the local electron density \citep[$\nu \propto \sqrt{n_e}$,][]{2010ARAandA..48..241B}.
These plasma emissions manifest as intense solar radio bursts, either from electrons accelerated at CME-driven shocks (Type~II bursts) or from beams of energetic electrons produced during solar flares (Type~III bursts).
Together, Type~II and Type~III radio bursts are a reliable tracer of transient mass motions, particle acceleration, and a diagnostic of coronal and wind properties for the Sun \citep{2017RvMPP...1....5M}.

There have been observations of events that may represent stellar analogs of Type~II and Type~III bursts \citep{2016ApJ...830...24C,2025Natur.647..603C,konijn_occurrence_2025}, but these observations have been conducted with ground-based radio telescopes observing at frequencies $\nu \gtrsim 50\,\mathrm{MHz}$.
Observations between approximately 10\,MHz and~50\,MHz suffer from phase fluctuations and absorption due to the Earth's ionosphere, and frequencies $\nu \lesssim 10\,\mathrm{MHz}$ do not penetrate the Earth’s ionosphere.
Because the plasma emission frequency depends on the ambient electron density---and because this density decreases with distance from the Sun---ground-based telescopes cannot track solar radio bursts farther than about~3$\,R_\sun$ from the Sun \citep{1998SoPh..183..165L}.
However, the acceleration of CME-driven solar energetic particles and the transition from the solar corona to the solar wind happens between about~5\,$R_\sun$ and~20\,$R_\sun$ \citep{2001JGR...10629219G,2005JGRA..11012S07G,2021PhRvL.127y5101K}.
Only so-called decametric-hectometric (DH) to kilometric radio bursts ($\nu \lesssim 15\,\mathrm{MHz}$) probe this distance range.

By extension, in order to track exospace weather, to determine if shocks and bulk mass motions are escaping to interplanetary space, and to constrain mass loss occurring via the stellar wind, access to distances farther than about~3$\,R_*$ is required.
For magnetically-active stars, plasma emissions occurring at comparable frequencies likely originate at $R \gtrsim 10\,R_*$ \citep{Davis2026}.

\begin{figure}[tb]
  \centering
  \makebox[0.66\textwidth][c]{%
    \includegraphics[width=0.66\textwidth]{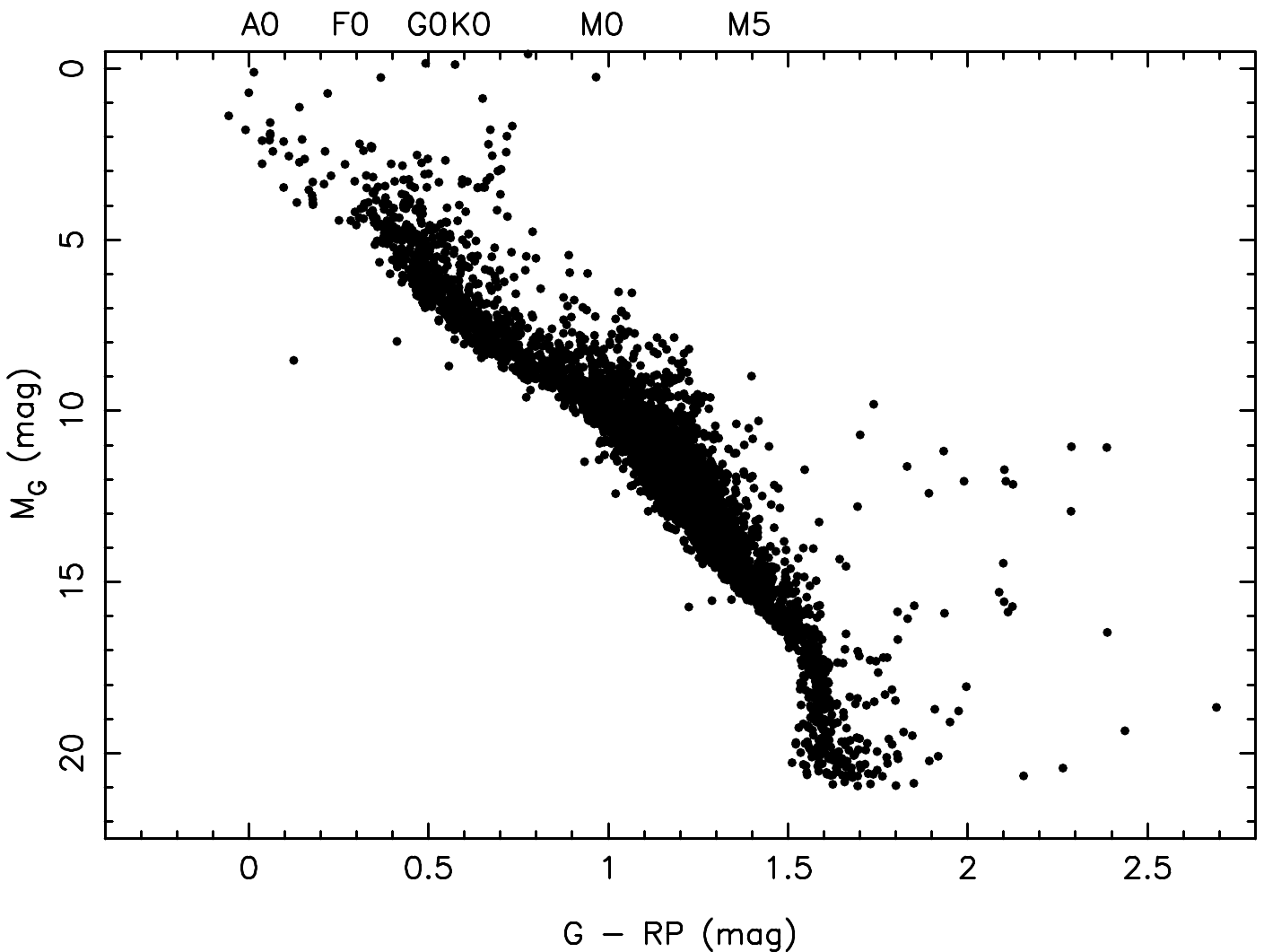}}\hfil%
  \parbox[b]{0.32\textwidth}{%
    \caption{There are over 5000 main-sequence stars within~25\,pc, of a wide range of spectral types, that the MegaWave Radio Surveyor could monitor to detect stellar radio bursts or magnetically-generated radio emission from exoplanets or both.
    (Adapted from the Fifth Catalogue of Nearby Stars [CNS5, \citealt{2023AandA...670A..19G}])}
      \vspace*{11ex}
  }
  \label{fig:nearby}  
\end{figure}

For stars of all spectral types, observing plasma emissions originating beyond $10\,R_*$ would provide strong evidence of energetic, transient mass escape into their interplanetary media and insight to steady mass loss at distances inaccessible to other observational methods.
The MegaWave Radio Surveyor would monitor at least 5000 stars within~25\,pc (Figure~2), of all spectral types hosting
hundreds of known exoplanets for radio bursts from
stellar CMEs and to provide diagnostics of their winds and for
exoplanetary radio emissions (next Science Question, \S\ref{sec:magnetic}).

\parbox[t]{0.95\textwidth}{%
\begin{description}
  \item[Science Objective] To determine the extent to which stars generate powerful transient stellar space weather events that eject material into their stellar winds and to determine properties of their winds
\end{description}
}

This investigation is synergistic with those of space physics
environments in other stellar and
planetary systems identified under
the ``New Environments: Exploring
Our Cosmic Neighborhood and
Beyond'' Theme in \textit{The Next
Decade of Discovery in Solar \& 
Space Physics} Decadal Survey.

\subsection{Decadal Science Question: What are the properties of individual planets, and which processes lead to planetary diversity?}\label{sec:magnetic}

Following the discovery of Jupiter's radio emission at~22\,MHz, it was proposed to be produced by Jupiter's planetary-scale magnetic field \citep{1955JGR....60..213B,1958JGR....63..807F,1969ARA&A...7..577C}.
Subsequent spacecraft investigations have revealed that the Earth, Mercury, Jupiter, Saturn, Uranus, and Neptune sustain internal dynamos that generate global magnetic fields.
Interactions between the solar wind and these magnetic fields generate intense radio emissions via the electron cyclotron maser instability \citep{z92,1999Ap&SS.264..401M}.
Jupiter's radio emission shows an additional signature caused by its moons \citep{1964Natur.203.1008B,1965Sci...148.1585D,1965Sci...148.1724L,2018A&A...618A..84Z}, raising the possibility of exomoon detection.

The existence of planetary magnetic fields and their properties reveal key information on the planetary  composition, heat transport mechanism, and dynamics \citep{KISS_MagneticPlanets,Driscoll2018,2025AREPS..53..305S}. This is because a dynamo must be formed in a region within the planet that is convective and the material is electrically conducting; rotation can also be important.
The location where the dynamo is generated affects the nature of the magnetic field. For example, 
Earth, Jupiter, and Saturn have largely dipolar magnetic fields due to magnetic generation near the planetary center, whereas the complex and multi-polar magnetic fields of Uranus and Neptune are thought to be generated further out, closer to the planetary ``surface"  (Figure~3). Characterizing the magnetic fields of planets can therefore be used to put limits on the planetary composition and the temperature profile where the field is generated. In return, this can also be used to constrain the long-term thermal evolution of the planet.  

\begin{figure}
  \centering
  \includegraphics[width=0.95\textwidth]{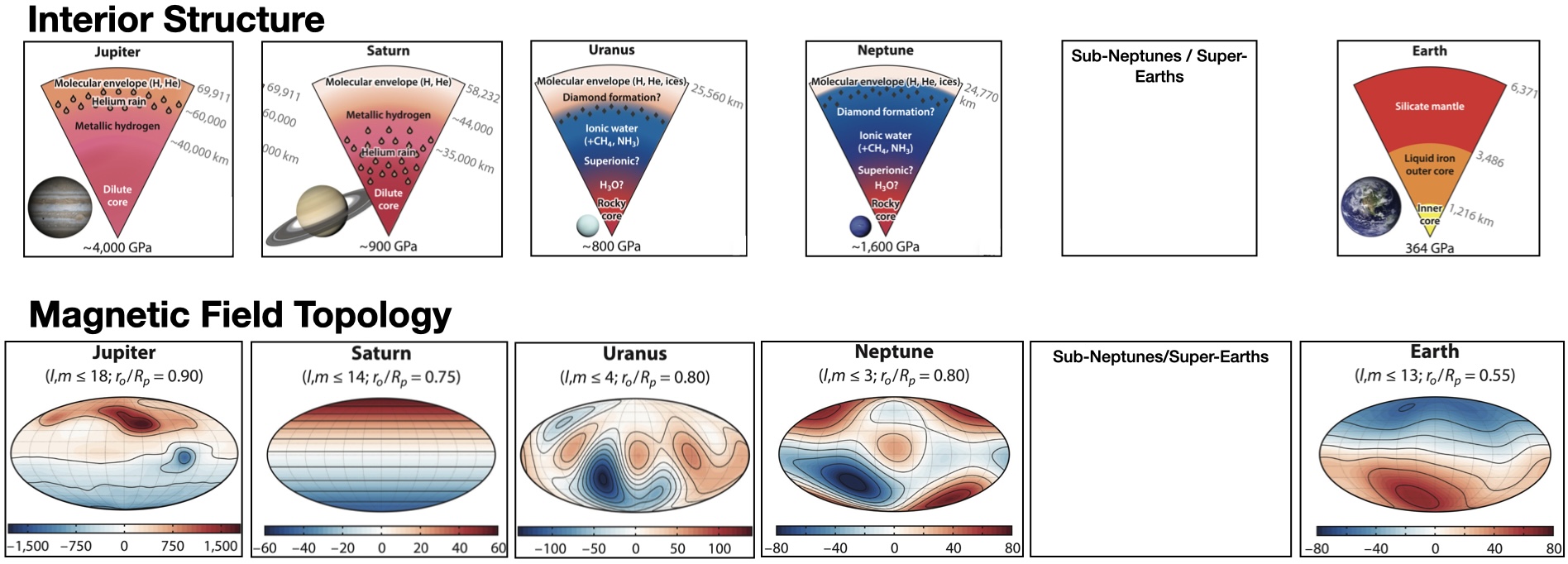}
  \caption{Magnetic fields probe planetary interior compositions and dynamics, but the small number of magnetized Solar System planets limit our understanding of dynamo processes and the extent to which Solar System planets are typical or outliers.  By detecting the magnetically-generated radio emission from exoplanets in the solar neighborhood, the MegaWave Radio Surveyor would improve our understanding of the interior structures of exoplanets and provide key information about the extent to which magnetic fields shield their atmospheres from particle erosion. (Adapted from \citealt{2025AREPS..53..305S}.)}
  \label{fig:interiors}  
\end{figure}

Observations at visible wavelengths of magnetic star-planet interactions in hot-Jupiter systems \citep{Shkolnik08,shkolnik2018b,strugarek2025} are consistent with expectations that the magnetic fields of Jupiter-mass planets are generated by metallic hydrogen, leading to strong and global magnetic fields \citep{Helled2020}.
New inferences of ultra-hot Jupiter magnetic fields from their atmospheric circulations indicate that their field strengths are consistent with those of Solar System planets \citep[and references within]{2026NatAs.tmp..116S}, but also that the strong stellar irradiation can, in some cases, suppress magnetic fields \citep{yadav2017ApJ...849L..12Y,2019NatAs...3.1128C}.
Clearly, inferred magnetic field strengths remain model-dependent.

The interior structures of intermediate-mass exoplanets, and particularly of sub-Neptunes and super-Earths, are even less constrained \citep[Figure~3;][]{Uranus}.
Not only are there no such planets in the Solar System, but their compositions and internal structures remain unknown.
In addition, the physical properties of materials in the relevant pressure-temperature conditions remain unknown \cite[e.g.,][and references therein]{Eberlein2025}. 
It is unclear whether the interiors of such planets are mostly adiabatic and largely convective, which should generate dynamos and planetary-scale magnetic fields, or consist of large regions where heat is transported by other mechanisms such as semi-convection,  conduction, and radiation.
In the latter case, the magnetic fields would be expected to be weaker and possible multi-polar  \citep{2013ApJ...768..156Y,2020RSPTA.37890479S,2020JGRE..12506124B,2022ApJ...938..131Z}.
Finally, the presence of basal magma oceans can supplement the magnetic fields driven by the core dynamos of these planets by nearly an order of magnitude --- an effect that can take place before their core dynamos even turn on, and last over a billion years into the lifetime of the planet  \citep{nakajima2026NatAs..10..248N}.

Detection of magnetically-generated radio emissions would provide an independent and complementary constraint for exoplanetary characterization, which could be combined with constraints from mass and radius measurements (and atmospheric characterizations), to constrain their internal structures, compositions, and heat transport, as the basic conditions to sustain a dynamo (convection, electrical conduction, and moderate rotation) must be met. 

The luminosities of wind-driven auroral emissions also directly diagnose to stellar wind properties \citep{saur2013A&A...552A.119S, 2018A&A...618A..84Z}, providing another means to assess space weather that is independent from stellar emissions (\S \ref{sec:exospace}) and new inputs for models of atmospheric loss in exoplanets. 

Magnetic fields often are described as ``shields'' for terrestrial planets (Figure~4; \citealt{2019MNRAS.485.3999M,2007SSRv..129..279D,2016GGG....17.1885F,2017ApJ...844L..13G,2019MNRAS.490...15O,HAMANO2025541}), and specifically for the Earth, protecting their secondary atmospheres and surfaces from high-energy charged particles (\S\ref{sec:exospace}) and loss of hydrogen.
Yet, in some cases, planetary magnetic fields can exacerbate atmospheric loss of certain species \citep{2019MNRAS.486.1283E,2019MNRAS.488.2108E,Driscoll2025}.
Planetary magnetic fields likely have a critical but complex role both for atmospheric shielding and contributing to planetary habitability, but, with only a small number of planetary magnetic fields to study in the Solar System, significant uncertainty remains.
Identifying the magnetic fields of terrestrial exoplanets in the solar neighborhood, particularly those to be observed with the \hbox{HWO}, will be a key factor in understanding their potential habitability.

\begin{figure}
  \centering
  \makebox[0.66\textwidth][c]{%
    \includegraphics[width=0.66\textwidth]{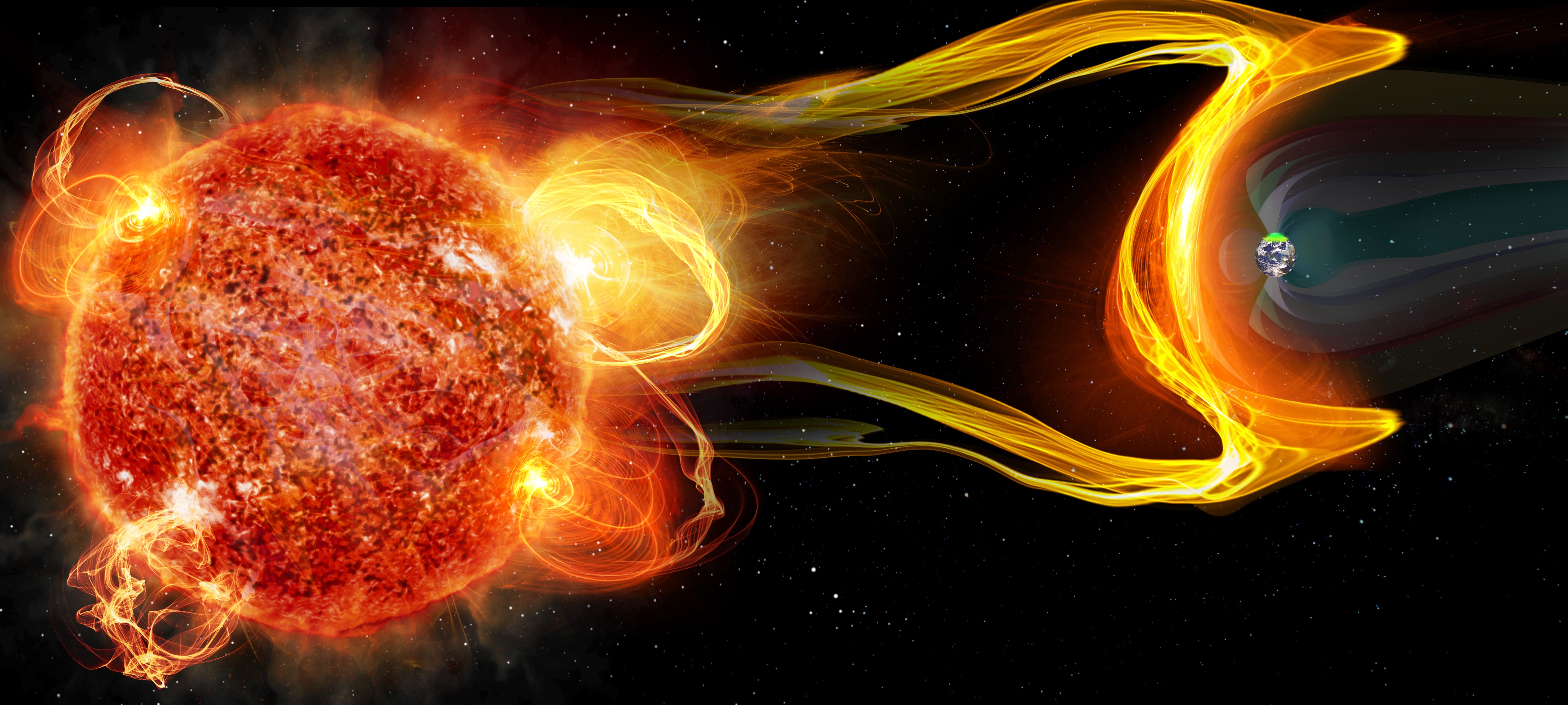}}\hfil%
  \parbox[b]{0.32\textwidth}{%
    \caption{The MegaWave Radio Surveyor would use magnetically-generated emissions from exoplanets to assess the role of magnetic fields as a contributor to habitability and as a probe for extrasolar planetary interiors. (Credit: Keck Institute for Space Studies/Chuck Carter)}}
  \label{fig:magnetic}  
\end{figure}

\parbox[t]{0.95\textwidth}{%
\begin{description}
 \item[Science Objectives]\phantom{blah}
    \begin{enumerate}
    \item To determine if exoplanet interiors support a more diverse
      range of magnetic fields than Solar System planets; and
    \item To determine if magnetic fields provide effective
      shielding of secondary atmospheres from escape
    \end{enumerate}  
\end{description}
}

The required measurement for these objectives is straightforward: The maximum radio frequency at which a planet emits via the electron cyclotron maser (ECM) instability is directly proportional to the magnetic field strength, $\nu_{\mathrm{ECM}} \simeq 2.8\,\mathrm{MHz}(B/1\,\mathrm{G})$ \citep{Griessmeier2018,2024arXiv240412348L,Zarka2025}.
Jupiter emits at frequencies up to approximately 35\,MHz, but the other magnetic planets in the Solar System have weaker magnetic fields and produce their most intense emissions at frequencies between about~0.3\,MHz and~1\,MHz.
Indeed, the Earth's own radio emission was not discovered until the advent of space-based radio instruments.
The electron cyclotron maser naturally produces circularly-polarized emission, often approaching 100\% polarization, making radio-emitting planets clearly distinguishable from other potential sources, especially at frequencies of tens of Megahertz.

This investigation is synergistic with the studies of ice giant
interiors and planetary
magnetospheres identified in the
\textit{Origins, Worlds, and Life}
Planetary Science \& Astrobiology
Decadal Survey (``What
processes influence the structure,
evolution, and dynamics of giant
planet interiors, atmospheres,
and magnetospheres?''), it addresses the
potential role of magnetic fields in contributing to a planet’s habitability as discussed in
the \textit{Exoplanet Science Strategy}, and it
would  address an Exoplanet Exploration Science Gap (``SCI-14:
Exoplanet interior structure and material properties'').

\subsection{Science Frontier: Uniquely probe the Universe’s evolution during the Dark Ages}\label{sec:darkages}

The ``Dark Ages'' is the time before the formation of the first stars,
at redshifts $1100 \gtrsim z \gtrsim 30$, and has been  identified as a science
discovery area in two  Decadal Surveys for Astronomy \& Astrophysics. Observations of hydrogen gas backlit by the Cosmic Microwave
Background (CMB) during the Dark Ages provides a unique probe of the Universe's
evolution during this era (Figure~5), enabling entirely new tests in cosmology and high-energy
physics.

\begin{figure}[bt]
  \makebox[0.66\textwidth][c]{%
    \includegraphics[width=0.66\textwidth]{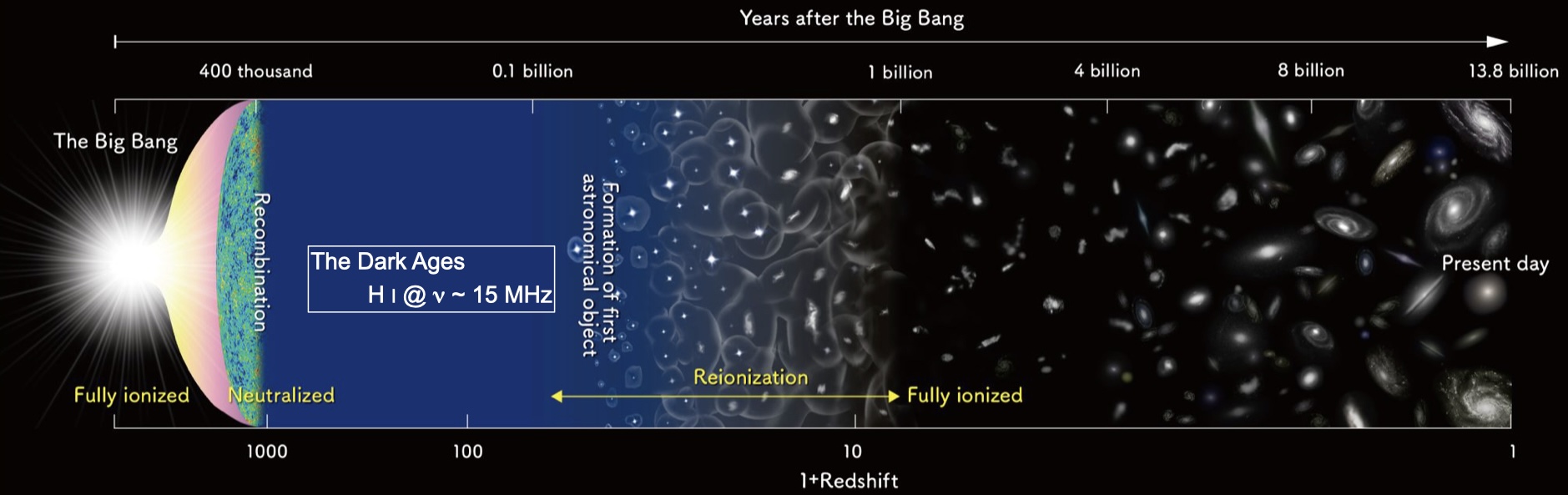}}\hfil%
  \parbox[b]{0.32\textwidth}{%
    \caption{The MegaWave Radio Surveyor would use the highly-redshifted hyperfine line of neutral hydrogen (\ion{H}{1}) as a unique probe of the Universe's Dark Ages.  (Credit: NAOJ)
      \label{fig:darkages}}}
\end{figure}

The quantum mechanics of the \ion{H}{1} atom produces
a hyperfine line at a (rest) frequency of~1420\,MHz ($\lambda \approx 21\mathrm{cm}$).
The basic physics has been presented by multiple authors \citep{1958PIRE...46..240F,1997ApJ...475..429M,2006PhR...433..181F,2007PhRvD..76h3005L,2008PhRvD..78j3511P,2012ApJ...760....4O,2012RPPh...75h6901P,2014PhRvD..89h3506A,2016JCAP...09..023C}:
as the Universe expands, the CMB (photon gas) and the \ion{H}{1} gas cool at different rates at $z \lesssim 200$.

A robust prediction of the $\Lambda$CDM model is that
there should be a sky-averaged (``global'' or monopolar) absorption signal from the \ion{H}{1} hyperfine line, strongest at observed frequencies of
about~15\,MHz, from the disequilibrium between the temperatures of the CMB and the
\ion{H}{1} gas during the Dark Ages.
This \ion{H}{1} signal is akin to the monopolar CMB signature detected by \cite{1965ApJ...142..419P}, albeit occurring in absorption rather than emission.

Further, due to cosmological redshift of this spectral line, the 3-D distribution of \ion{H}{1} during the Dark Ages could be mapped tomographically, with observing frequency encoding radial distance, thus combining the best aspects of previous CMB missions (\hbox{WMAP}, Planck) and intensity mapping surveys such as SPHEREx.
Because of the large redshifts involved, a
considerable fraction of the observable Universe is accessible, expanding upon the
surface of last scattering probed by CMB missions (and without the ``Silk'' or diffusion damping that suppresses small-scale CMB fluctuations).
The Universe was in a uniquely simple evolution state during the Dark Ages, as stars and galaxies had not yet formed, so most potential astrophysical uncertainties and biases, which experiments like SPHEREx must correct, are not present.
The combination of large volume, small-scale
measurements, and a relatively simple state of the Universe provides a statistical power up to~$10^5 \times$ higher than the \hbox{CMB}, yielding unprecedentedly precise cosmological constraints \citep{2004PhRvL..92u1301L,2008PhRvD..78b3529M,2021MNRAS.501.2627C,2026arXiv260113053S}.

A number of authors have illustrated the power of the Dark Ages \ion{H}{1} signal, including ultimate, cosmic-variance limited constraints on the inflationary power spectrum~\citep{Munoz:2015eqa,Meerburg:2016zdz}, new limits on neutrino masses~\citep{Pritchard:2008wy}, and constraints on the nature of dark matter at small scales~\citep{Munoz:2016owz}.
Further, any injection of energy into the \ion{H}{1}
gas or inhomogeneities will
distort the expected power spectra, whether due to decays of particles beyond the
Standard Model, effects of primordial magnetic fields, shocks due to relative motions
between hydrogen and dark matter, accretion of gas onto primordial black holes, spatial
inhomogeneities in the \hbox{CMB}, or other exotic physics \citep{2002PhRvD..65l3517D,2004PhRvD..70d3502C,2009ApJ...692..236S,Tashiro:2014tsa,Munoz:2015bca,2018JCAP...05..017B,2018PhRvD..98j3505A,2023NatAs...7.1025M,Libanore:2025ack,2026MNRAS.548ag581P,2026arXiv260822319A}.
The Dark Ages represent the final and most definitive frontier to measure the cosmological model of our Universe. 

\parbox[t]{0.95\textwidth}{%
  \begin{description}
   \item[Science Objectives]\phantom{blah}
     \begin{enumerate}
     \item To test novel physics beyond the standard cosmological
       model through its effects on the thermal history of the
       intergalactic medium in the Dark Ages; and 
     \item To test concordance of structure formation in the Dark Ages
       with inflationary era predictions.
     \end{enumerate}
  \end{description}
  }

The MegaWave Radio Surveyor would focus initially on the detection and characterization of the global Dark Ages \ion{H}{1} signal, as it is the brighter of the two observational signatures.
On the longer term, the focus would shift to tomography, measuring the 3-D angular power spectrum of the \ion{H}{1} fluctuations.

\subsection{Decadal Science Question: How do gas, metals, and dust flow into, through, and out of galaxies?\linebreak
Decadal Science Question: How do the histories of galaxies and their dark matter halos shape their observable properties?}\label{sec:crs}

Multi-wavelength observations increasingly indicate that cosmic rays (CRs) and magnetic fields
play a significant role in the cosmic web, contributing to the structures of galaxies and clusters of galaxies, as a fraction of the energy released during structure formation is channeled into turbulence, magnetic field amplification,
and acceleration of CR electrons \citep{heesen_radio_2021,2023Univ....9..319W}.

Galactic winds and outflows, whether driven by black hole accretion or vigorous stellar feedback, should contain a CR component, the magnitude of which could affect their multi-phase circumgalactic media \citep[\hbox{CGM},][]{2020MNRAS.492.3465H,2022ApJ...935...69B,2025arXiv251014908S,ponnada_time-dependent_2026,2026arXiv260706744W}.
For instance, the magnitude of non-thermal pressure contributed by CRs and magnetic fields may explain puzzling absorption line ratios observed in the CGM of various galaxies along lines of sight toward QSOs \citep{Ji2020,butsky_impact_2020,thomas_why_2025,lu_constraining_2026}, as well as the extended soft X-ray emission observed by eROSITA \citep{hopkins_cosmic_2025} in conjunction with the depressed total CGM thermal pressure as detected via the thermal Sunyaev-Zel'dovich (SZ) effect \citep{2026ApJ...997L..13P}.

Galaxy clusters, the largest gravitationally-bound structures in the Universe, nodes in the cosmic web, serve as precise
tracers of cosmic structure formation and evolution \citep{2012ARA&A..50..353K}.
Galaxy clusters form and grow through continuous matter accretion from the surrounding cosmic web and via energetic merger events, with the most energetic mergers releasing up to~$10^{62}\,\mathrm{ergs}$.
The basic physics dictates that shocks generated during these processes convert gravitational energy into thermal energy of the intracluster medium (ICM).
A significant fraction of this energy is also channeled into turbulence, magnetic field amplification, and the acceleration of CRs \citep{2008Sci...320..909R,2012MNRAS.421.3375V,2014IJMPD..2330007B,2019SSRv..215...16V}.

The shocks resulting from cosmic structure formation can be revealed through structures such as radio relics (Figure~6), thought to originate from shock-accelerated electrons in the \hbox{ICM}.
However, how the energy released during structure formation is partitioned between thermal plasma, magnetic fields, and relativistic particles across galaxy clusters and the cosmic web remains uncertain.
In some cases, extended radio emission is observed between clusters well before their merger, forming ``radio bridges'' that are thought to trace shocks and turbulence in the pre-merging region \citep{2019Sci...364..981G,2020PhRvL.124e1101B}.
A particular puzzle results from comparisons between predicted and observed radio luminosities of extended radio emission in various clusters of galaxies.  
The shocks in clusters of galaxies have Mach numbers ${\cal M} \sim 3$, for which the standard diffusive shock acceleration (DSA) process predicts relatively low efficiency for CR acceleration.
The radio luminosities of some radio relics, however, require an apparently unphysical CR acceleration efficiency,  exceeding 100\% under standard DSA assumptions \citep[e.g.,][]{2020A&A...634A..64B}.
These results suggest that additional processes, such as the reacceleration of pre-existing ``fossil'' (low-energy) CR electrons, may contribute to the observed radio emission.

\begin{figure}[bt]
  \makebox[0.60\textwidth][c]{%
    \includegraphics[width=0.60\textwidth]{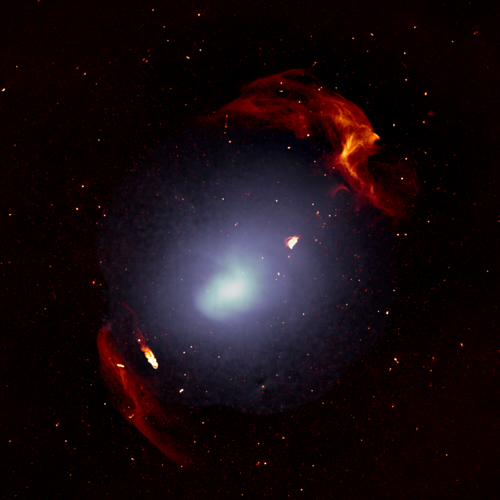}}\hfil%
  \parbox[b]{0.35\textwidth}{%
    \caption{Radio relics, as illustrated (in orange) in Abell~3667, trace shocks associated with cosmic structure formation, representing an interface between the thermal (X-ray emitting [blue]) gas in the cluster's core and the larger cosmic web.  Both the intracluster media (ICM) of clusters of galaxies and the circumgalactic media (CGM) of galaxies should be pervaded by low-energy cosmic ray electrons.  The MegaWave Radio Surveyor would observe the synchrotron emission to trace processes associated with cosmic structure formation and galaxy evolution.  (Credit: F.~de~Gasperin)
  }}
  \label{fig:shocks}  
\end{figure}

As galaxies move in groups, shocks should form at the interface between the galaxies' CGM and the intragroup media.
These termination shocks could accelerate particles (protons or ions) to energies $E \sim 10^{16}\,\mathrm{eV}$ \citep{2017ApJ...835...72B,2018ApJ...859...63M}.
The source or sources that produce CRs in the energy range between about $3 \times 10^{15}\,\mathrm{eV}$ (the ``knee'' of the CR energy spectrum) and $10^{18}\,\mathrm{eV}$ (the ``ankle'') is uncertain.
If galactic termination shocks amplify magnetic fields and accelerate CRs, both protons/ion and electrons, detecting radio emission associated with them would clarify relevant source population in this energy range of the CR spectrum.

\parbox[t]{0.95\textwidth}{%
  \begin{description}
  \item[Science Objective] To determine the significance of cosmic rays
    and magnetic fields, including their non-thermal pressure contributions, in intracluster and circumgalactic media
  \end{description}
  }

In these environments, CR electrons with energies $E \lesssim 1\,\mathrm{GeV}$ in environments with sub-$\mu$G to~$\mu$G magnetic fields \citep{2021MNRAS.502.2518S,2022SciA....8.7623B,ponnada_magnetic_2022,ramesh_circumgalactic_2023} emit synchrotron radiation at frequencies $\nu \lesssim 15\,\mathrm{MHz}$ and have long residence times, enabling them to serve as tracers of energy injection and the history of the cosmic structure formation and galaxy evolution.
Such low-frequency observations, in tandem with recent works suggesting extended soft X-ray emission from $E \lesssim 1\,\mathrm{GeV}$ CR electrons, would present unprecedented constraints on the non-thermal physics of magnetic fields and CRs in galactic halos.

Finally, while requiring additional development of the science objectives and requirements, it is likely that the MegaWave Radio Surveyor could address questions related to the rate of cosmic ray ionization within molecular clouds within the Milky Way \citep{2009A&A...501..619P,2020SSRv..216...29P}.

\clearpage

\begin{deluxetable}{|c|c|c|c|}
  \tablewidth{0pc}
  \tabletypesize{\footnotesize}
  \tablecaption{Traceability Table\label{tab:stm}}
  \tablehead{
    \colhead{\textbf{Science Objectives}} & \colhead{\textbf{Physical Parameters}} &
    \colhead{\textbf{Observables}} & \colhead{\textbf{Potential Challenges}}
  }
  \startdata
   \parbox[c]{0.22\textwidth}{%
   \textbf{Exospace Weather} (\S\ref{sec:exospace})
     To determine the extent to which stars generate powerful transient
     stellar space weather events that eject material into their stellar
     winds and to determine properties of their winds}
   &
   \parbox[c]{0.22\textwidth}{%
     Energetic particles at distances $\gtrsim 10\,R_*$, where they
     excite plasma emissions that radiate at frequencies based on the local
     stellar wind densities}
   &
   \parbox[c]{0.22\textwidth}{%
     Intensity as a function of time and frequency at decametric-hectometric (DH) frequencies below
     15\,MHz, measured for a sufficient sample of stars}
   &
   \parbox[c]{0.22\textwidth}{%
     \begin{itemize}
      \item Importance of frequency range has been realized in past two decades
      \item Capabilities for space-based interferometers with sufficient
        performance have not existed previously
     \end{itemize}
   }
    \\
    \noalign{\hrule}
    \parbox[c]{0.22\textwidth}{%
    \vspace*{1ex}
    \textbf{Magnetic Planets} (\S\ref{sec:magnetic})\hfil%
      \begin{enumerate}
      \item To determine if exoplanet interiors support a more diverse
        range of magnetic fields than Solar System planets; and
      \item To determine if magnetic fields provide effective
        shielding for secondary atmospheres
      \end{enumerate}
      }
    &
    \parbox[c]{0.22\textwidth}{%
      Planetary magnetic field strength~$B$, which determines the electron
      cyclotron maser emission frequency $2.8\,\mathrm{MHz}(B/1\,\mathrm{G})$}
    &
    \parbox[c]{0.22\textwidth}{%
      Total and circular polarized intensity over the frequency range of at least 0.5\,MHz to~30\,MHz, measured
      for a sufficient sample of exoplanets}
    &
    \parbox[c]{0.22\textwidth}{%
      Capabilities for space-based interferometers with sufficient
      performance have not existed previously}
    \\
    \noalign{\hrule}
    \parbox[c]{0.22\textwidth}{%
    \vspace*{1ex}
    \textbf{Dark Ages} (\S\ref{sec:darkages})\hfil%
      \begin{enumerate}
      \item To test novel physics beyond the standard cosmological
        model through its effects on the thermal history of the
        intergalactic medium in the Dark Ages; and 
      \item To test concordance of structure formation in the Dark Ages
        with inflationary era predictions.
      \end{enumerate}
      }
    &
    \parbox[c]{0.22\textwidth}{%
      At redshifts $150 \gtrsim z \gtrsim 30$,
      \begin{enumerate}
       \item Temperature, and
       \item Density fluctuations
      \end{enumerate}
      of neutral hydrogen gas}      
    &
    \parbox[c]{0.22\textwidth}{%
      At frequencies $7\,\mathrm{MHz} \lesssim \nu \lesssim 45\,\mathrm{MHz}$
      \begin{enumerate}
      \item Amplitude and shape of global {H}\,\textsc{i} signal
      \item Amplitude and shape of {H}\,\textsc{i} power spectrum
      \end{enumerate}
      }
    &
    \parbox[c]{0.22\textwidth}{%
      \begin{tabitemize}
       \item Receiver stability and calibration knowledge is at or beyond
        state of the art
       \item Capabilities for space-based interferometers with sufficient
        performance have not existed previously
      \end{tabitemize}
      }
    \\
    \noalign{\hrule}
    \parbox[c]{0.22\textwidth}{%
    \vspace*{1ex}
    \textbf{Low-Energy Cosmic Rays (\S\ref{sec:crs})}\hfil\\
      To determine significance of cosmic rays and magnetic fields,
      including their non-thermal pressure contributions, in
      intracluster and circumgalactic media}
    &
    \parbox[c]{0.22\textwidth}{%
      Synchrotron emissivity as function of position within intracluster
      and circumgalactic media}
    &
    \parbox[c]{0.22\textwidth}{%
      Surface brightnesses of intracluster and circumgalactic media at
      frequencies below 15\,MHz}
    &
    \parbox[c]{0.22\textwidth}{%
      Capabilities for space-based interferometers with sufficient
      performance have not existed}
  \enddata  
\end{deluxetable}

\clearpage

\section{Instrument Description}\label{sec:instrument}

The MegaWave Radio Surveyor, a space-based \textbf{Interferometer} (Figure~7) of unprecedented scale composed of hundreds of mass-produced smallsats, would address the science goals outlined above and would serve as a Formative Era concept in the \textit{Enduring Quests, Daring Visions} roadmap.
It also would enable NASA to develop a new paradigm for space observatories.
As a distributed aperture, the MegaWave Radio Surveyor concept blends traditional divisions between the ``instrument'' and the ``mission'' (\S\ref{sec:mission}), thereby pioneering the methodologies and operational approaches for future missions and expanding NASA's capabilities beyond the traditional single-spacecraft model.

The MegaWave Radio Surveyor would obtain images by measuring the spatial coherence function, commonly termed the ``visibility function,'' which is the second moment of the electric field, $V_{ij} = \langle E_iE^*_j\rangle$ from antennas~$i$ and~$j$.
Standard approaches exist to invert the visibility function to produce the sky brightness or analyze the visibility function directly \citep{tms3,2026NatAs..10..410T}.

\begin{figure}[h]
 \centering
 \includegraphics[width=0.95\textwidth]{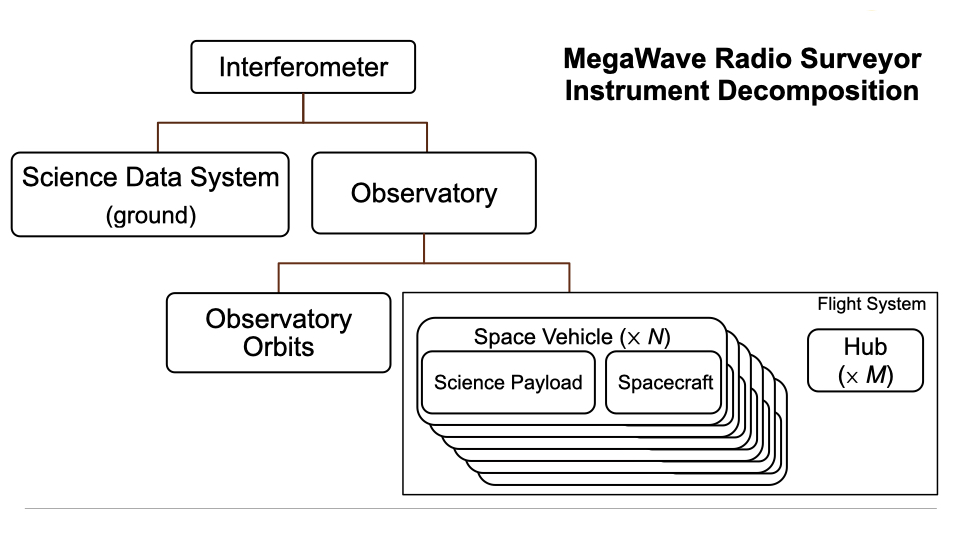}
 \vspace*{-3ex}
 \caption{A hierarchical decomposition of the MegaWave Radio Surveyor is based on system engineering from NASA’s Sun Radio Interferometer Space Experiment \citep[\hbox{SunRISE},][]{SunRISE}.}
  \label{fig:instrument}  
\end{figure}

The flight system consists of an \textbf{Observatory},
which itself is composed of~$N$ identical \textbf{Space Vehicles} and~$M$~\textbf{Hubs}; this architecture also is planned for NASA's HelioSwarm mission \citep{2023SSRv..219...74K}, albeit on a smaller scale, and a similar architecture has
been developed for the Great Observatory for Long Wavelengths (GO-LoW) concept \citep{GOLoW}. Sensitivity and
angular resolution requirements determine both the number of Space Vehicles and their orbits.
The \textbf{Hubs} would distribute time/frequency reference signals and could serve as communication
relays. A \textbf{Space Vehicle} consists of a \textbf{Science Payload} and a \textbf{Spacecraft}. The \textbf{Science Payload}
is responsible for receiving and recording the astronomical radio waves. The \textbf{Spacecraft} hosts
the Science Payload and carries avionics, propulsion, telecommunications, power, and related
sub-systems. The ground-based \textbf{Science Data System} produces higher-level data products to
meet the Science Objectives, which would be stored in an Astrophysics
Archive.

There have been a number of specific ``point designs'' developed previously for space-based interferometers observing at low radio frequencies.
However, consistent with the intent of NASA/Astrophysics' Strategic Technology \& Research Accelerator (ASTRA) Initiative focus on mission concepts at Concept Maturity Levels 2--3 \citep{CML} and on-going (disruptive) changes in the space mission landscape, we highlight a number of trade studies and enabling capabilities that warrant (re-)examination: 
\begin{itemize}
 \item \textbf{Antenna topologies} and the fraction of Payloads with
   \textbf{augmented calibration requirements};
 \item  Number of Hubs, which could depend upon availability of
   \textbf{commercial relay satellites} and the accessibility of
   commercial lunar relay services that will be available through the planned Lunar Communications Relay \& Navigation Systems (LCRNS);
 \item \textbf{Science return analysis} to determine the value of
   \textbf{parallel observations}, e.g., by dividing the
   interferometer into ``sub-arrays,'' with one set of Space Vehicles
   tasked for observing one target and another set observing a second
   target, with the number of sub-arrays and Space Vehicles per
   sub-array adjusted depending upon the science targets being
   observed;
 \item Balance between on-board (Spacecraft or Hub) and ground
   processing based on the \textbf{transmitted data volumes} and the
   precision of \textbf{space-based orbit determination};
 \item Extent to which U.{}S.\ space industries can be leveraged for
   \textbf{mass manufacturing} in order to obtain economies of scale
   and deliver a large number of Space Vehicles;
 \item Availability of \textbf{high-speed space communications}, e.g., the maturity of deep-space laser communications capabilities, and \textbf{advanced space computational hardware}, for real-time orbit determination and initial data processing; and
 \item Improvements and refinement of algorithms for \textbf{precision cosmology at decameter wavelengths}, e.g. calibration and filtering methods for separating the signal from instrumental effects, radio-frequency interference, and astrophysical foregrounds.
\end{itemize}

\section{Mission Implementation}\label{sec:mission}

The MegaWave Radio Surveyor concept is now technically feasible to design, enabled by developments in the U.{}S.\ space industry including the standardization of small spacecraft, advances in onboard processing and communications, commercial spacecraft mass production, and increasing heritage in distributed small spacecraft missions.
The available trade space allows for leveraging elements of the Artemis/Moon-to-Mars program, including launches to cislunar space and communications and navigation capabilities.

The MegaWave Radio Surveyor concept leverages
experience with the development of the Sun Radio Interferometer Space Experiment \citep[\hbox{SunRISE},][]{SunRISE}, the development and operations of the Star-Planet Activity Research CubeSat \citep[\hbox{SPARCS},][]{shkolnik25}, the development of the Lunar Surface Electromagnetics Explorer at Night \citep[LuSEE-Night,][]{2023arXiv230110345B,10906958,10907089}, and decades of experience with ground-based radio observatories, such as the 74\,MHz system on the Very Large Array \citep{2007ApJS..172..686K}, the Low Frequency Array \citep[\hbox{LOFAR},][]{LOFAR}, and the Long Wavelength Array \citep[\hbox{LWA},][]{2012JAI.....150004T,2025JAI....1450003T}.

Key trade studies or optimizations for the MegaWave Radio Surveyor, including aspects that would leverage the Artemis/Moon-to-Mars program, include
\begin{description}
 \item[Orbit]%
   The Science Objectives require space-based observations, as the relevant frequencies are
   distorted or entirely blocked by the Earth's ionosphere. The orbits of the Space Vehicles within
   the Observatory and of the Observatory itself have multiple options.
   \begin{enumerate}
   \item Space Vehicle orbits could be designed to provide a wide and
     nearly uniform range of separations or could favor specific
     separations, based on optimization for the Science Objectives. An
     additional science return analysis could consider the science
     value of reconfiguring the orbits.
   \item The Observatory itself could be positioned at the Earth-Moon Lagrange point L1, L2, L4, or L5, or Sun-Earth L2.
   \end{enumerate}
   Dark Ages \ion{H}{1} observations are facilitated by having a stable environment for the Science
   Payload, as experience with ground-based instruments has shown the importance of a
   well-modeled and highly stable instrument for achieving the dynamic range required for
   cosmological studies. Any of these orbits would provide nearly constant solar illumination for
   the Space Vehicles/Science Payloads. A low level of radio interference is desirable, but over a
   significant fraction of the required frequency range, the Earth's ionosphere provides partial or
   complete blockage of ground-based transmitters. Trade space parameters include at least orbital
   stability and stationkeeping budget, data volumes to be transmitted, mission operations, and the
   potential for radio interference from cislunar spacecraft.

 \item[Launch and Serviceability-Scalability]%
   The Space Vehicles and Hubs could be stacked into a
   single launch, as has been standard for previous Astrophysics missions. Alternatively, an initial
   launch could deploy a notional capability that would be expanded with additional launches.
   Interferometers are naturally scalable allowing additional Space Vehicles to be added over time,
   and this approach is a new perspective for serviceability in which the Observatory is replenished
   by replacing and adding Space Vehicles. This approach would leverage increased access to
   space, including potentially launches associated with the
   Artemis/Moon-to-Mars program.
 
 \item[Position, Navigation, \& Timing]%
   In order to combine signals coherently from the Science Payloads, they must have stable frequency/time references and the Space Vehicle orbits must be determined to a precision on the order of~1\,m.
   Options could include using the Lunar Communications Relay \& Navigation Systems (LCRNS) directly or having one or more Hubs broadcast a common frequency reference to all Space Vehicles.
   A hybrid option would be that the Hub or Hubs that provide the frequency reference signals to the Space Vehicles is modeled on or adapted from an LCRNS spacecraft.
   
 \item[Manufacturing Processes and Mission Operations]%
   The required number of Space Vehicles is larger than has been standard for the Astrophysics Division (or other Divisions within the Science Mission Directorate), but the number of Space Vehicles is comparable to or (much) fewer than those already achieved by multiple commercial entities, though those spacecraft have operated only in low-Earth orbit (LEO).
   Multiple companies are advertising the capability to produce hundreds of spacecraft per year, and assessing how to leverage demonstrated industrial and commercial capabilities for science missions would be enabling.
   Specifically, developing a testing methodology derived from commercial sector experience of satellite mass production, which allows for some small failure rate of individual spacecraft, will be essential to controlling costs and schedule for a constellation telescope.

    This incorporation from the commercial sector would have to flow through all stages of the mission design, from assembly and integration to test to operations.  
    In particular, new approaches to mission operations would be required given light travel time latencies and large number of vehicles.
    The balance between on-orbit processing, whether in Space Vehicles or in a Hub, and ground-based processing is a key trade, dependent in part upon communication capabilities.
    Infusing artificial intelligence/machine learning (AI/ML) capabilities may represent an opportunity or even be required for mission operations.
\end{description}

Potential international partnerships or contributions in implementing the MegaWave Radio Surveyor could include the following:
\begin{description}
\item[Europe]%
  Multiple individuals with potential scientific interests and multiple institutes with potential technological contributions to both the flight hardware and the ground-based post-processing, most notably from the operations of the Low Frequency Array (LOFAR) and post-processing of the data generated by it; specific contributions could range from providing the launch vehicle to specific satellite subsystems to a fraction of the satellites required for the total array to portions of the software base for ground-based processing.

\item[United Kingdom]%
  Multiple individuals with potential scientific interests and multiple institutes with potential technological contributions; contributions could range from providing satellite subsystems to supplying a fraction of the satellites required for the total array.

\item[India]%
    There are multiple individuals interested in the potential science case and having broad expertise in the domain of low radio frequency astronomy, including hardware development, analysis, and inference.  There are ongoing experiments for the highly-redshifted hyperfine \ion{H}{1} line and a proposed experiment for cosmic dawn studies in lunar orbit.  An interface with Indian Space Research Organization (ISRO) may be explored, in particular through participating academic institutions.
\end{description}

\begin{acknowledgments}
We thank R.~Amini and A.~Nash for helpful comments on the Instrument
Description and Mission Implementation sections.
The Dark Ages graphic on the cover image is adapted from an SDSS image generated by C.~Blake and S.~Moorfield.
The cover image itself was generated with the use of \hbox{ChatGPT}.
The GO-LoW team has been funded by NIAC grants 80NSSC23K0585 and 80NSSC24K1236.
This research has made use of the Astrophysics Data System, funded by NASA under Cooperative Agreement 80NSSC21M0056.
Basic research at NRL is funded by 6.1~Base programs.
Part of this research was carried out at the Jet Propulsion Laboratory, California Institute of Technology, under a contract with the National Aeronautics and Space Administration.
\end{acknowledgments}

\clearpage

\bibliographystyle{aasjournalv7}
\bibliography{MegaWaveRadioSurveyor}{}

\end{document}